\documentclass[journal,twoside,web]{ieeecolor}
\usepackage{jsen}
\usepackage{cite}
\usepackage{amsmath,amssymb,amsfonts}
\usepackage{algorithmic}
\usepackage{graphicx}
\usepackage{textcomp}
\usepackage{url}
\usepackage{float}
\usepackage[position=bottom]{subfig}
\usepackage[draft=true]{hyperref}
\hypersetup{final}
\usepackage{wrapfig}
\usepackage{amsmath,amssymb,amsfonts}
\usepackage[noabbrev]{cleveref}      
\usepackage{algorithmic}
\usepackage{textcomp}
\usepackage{booktabs}
\usepackage[T1]{fontenc}

\def\BibTeX{{\rm B\kern-.05em{\sc i\kern-.025em b}\kern-.08em
    T\kern-.1667em\lower.7ex\hbox{E}\kern-.125emX}}
\definecolor{abstractbg}{rgb}{0.89804,0.94510,0.83137}
\begin{document}
\title{Revealing Preference in Popular Music Through \\Familiarity and Brain Response}
\author{Soravitt~Sangnark,
        Phairot~Autthasan,~\IEEEmembership{Student Member,~IEEE},
        Puntawat~Ponglertnapakorn,
        Phudit~Chalekarn,
        Thapanun~Sudhawiyangkul,
        Manatsanan~Trakulruangroj,
        Sarita~Songsermsawad,
        Rawin Assabumrungrat,
        Supalak~Amplod,
        Kajornvut~Ounjai$^{*}$,
        and~Theerawit~Wilaiprasitporn$^{*}$,~\IEEEmembership{Member,~IEEE}
\thanks{This work was supported by PTT Public Company Limited, The SCB Public Company Limited, Thailand Science Research and Innovation (SRI62W1501) and Office of National Higher Education Science Research and Innovation Policy Council (C10F630057).} 
\thanks{S. Sangnark, P. Autthasan, P. Chalekarn, T. Sudhawiyangkul, S. Amplod, and T. Wilaiprasitporn are parts of Bio-inspired Robotics and Neural Engineering Lab, School of Information Science and Technology, Vidyasirimedhi Institute of Science \& Technology, Rayong, Thailand. {\tt\small ($^{*}$corresponding authors: theerawit.w at vistec.ac.th, kajornvut.oun@kmutt.ac.th)}.}
\thanks{P. Ponglertnapakorn is with Vision and Learning Lab, School of Information Science and Technology, Vidyasirimedhi Institute of Science \& Technology, Rayong, Thailand.}
\thanks{S. Songsermsawad is with School of Energy Science and Engineering, Vidyasirimedhi Institute of Science \& Technology, Rayong, Thailand.}
\thanks{K. Ounjai is with Biological Engineering program, Faculty of Engineering. He also is a part of Neuroscience Center for Research and Innovation, Learning Institute at King Mongkut's University of Technology Thonburi, Thailand.}
\thanks{M. Trakulruangroj is with Faculty of Electrical Engineering, Sirindhorn International Institute of Technology, Thammasat University, Thailand.}
\thanks{R. Assabumrungrat is with School of Engineering, Tohoku University, Japan.}
}

\IEEEtitleabstractindextext{%
\fcolorbox{abstractbg}{abstractbg}{%
\begin{minipage}{\textwidth}%
\begin{wrapfigure}[14]{r}{3.10in}%
\includegraphics[width=2.90in]{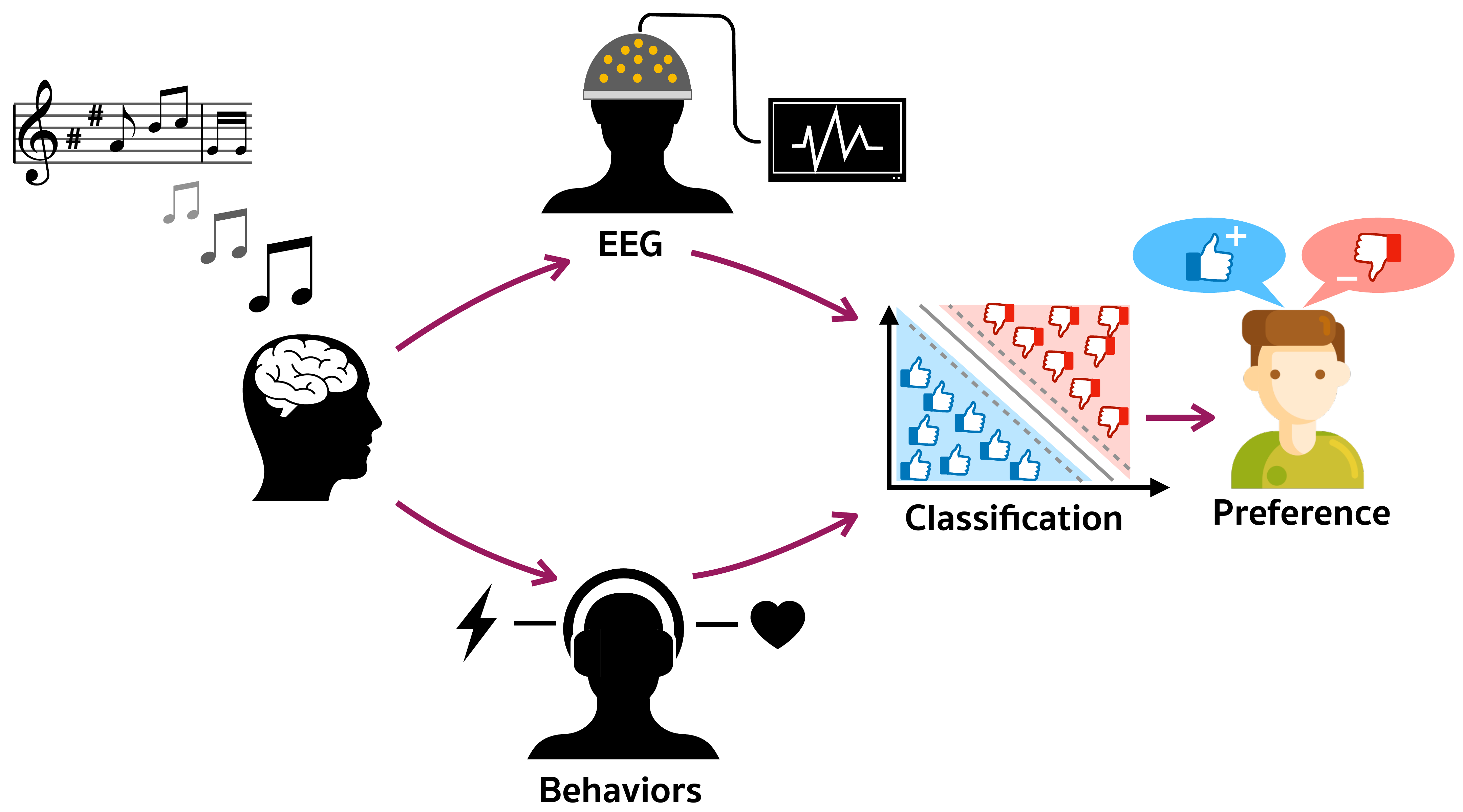}%
\end{wrapfigure}%
\begin{abstract}

Music preference was reported as a factor, which could elicit innermost music emotion, entailing accurate ground-truth data and music therapy efficiency. This study executes statistical analysis to investigate the distinction of music preference through familiarity scores, response times (response rates), and brain response (EEG). Twenty participants did self-assessment after listening to two types of popular music's chorus section: music without lyrics (Melody) and music with lyrics (Song). \textcolor{red}{We then conduct a music preference classification using a support vector machine, random forest, and k-nearest neighbors with the familiarity scores, the response rates, and EEG as the feature vectors. The statistical analysis and F1-score of EEG are congruent, which is the brain's right side outperformed its left side in classification performance.} Finally, these behavioral and brain studies support that preference, familiarity, and response rates can contribute to the music emotion experiment's design to understand music, emotion, and listener. Not only to the music industry, the biomedical and healthcare industry can also exploit this experiment to collect data from patients to improve the efficiency of healing by music.
\end{abstract}

\end{minipage}}}

\maketitle

\section{Introduction}
\IEEEPARstart{M}{usic} \textcolor{black}{enjoyment aesthetically roots in all human cultures \cite{kohut_1}. As the music database continuously grows, an effective method to organize music information to meet the audience's preference is needed. Music emotion recognition (MER) is such a method that has been continuously received attention from researchers \cite{yang_2, panda_30, berardinis_7}. However, music emotion depends on various factors. As a consequence, data reliability and prediction remains a matter of concern.}

\textcolor{black}{
In previous decades, several researchers have tried to investigate which factors have influenced music emotion. Aljanaki \textit{et al.} found that the self-reported emotional music scale was influenced by music preference \cite{aljanaki_7, aljanaki_19}. Naser and Saha found that preference influenced the performance of music emotion recognition \cite{naser_11}. Besides the preference, familiarity would be one of the most influential factors in music emotion study \cite{warrenburg_15}.}
\textcolor{black}{
Pereira \textit{et al.} stated that familiarity was a significant factor in determining listeners' emotional engagement \cite{pereira_6}. They also found a psychological phenomenon as \textit{mere exposure effect}, which defined as people tend to be more favored due to they were more familiar. The psychological result was similar to \cite{meyer_48, peretz_47}. Interestingly, familiarity and preference were proved to have a robust positive correlation \cite{north_20, hamlen_20}.}
\textcolor{black}
{Liljeström \textit{et al.} found that self-chosen music, which was more familiar and preferred, evoked more innermost emotion than randomly sampled music \cite{Liljes_21}. 
However, even though various studies showed that music preference and familiarity were essential, only a few music emotion studies focused on both of the factors.}



\textcolor{black}{
Well-known emotion datasets, such as DEAP \cite{koelstra_8}, ASCERTAIN \cite{subramanian_15}, and AMIGOS \cite{correa_21}, had routinely collected familiarity and music preference. However, a few studies applied these factors. For example, DEAP \cite{koelstra_8} was more utilized to investigate electroencephalography (EEG) for classifying emotion \cite{tao_17,cheng_22, gao_23} than preference \cite{aldayel_12} and familiarity \cite{sanggarini_9}. Moreover, although the usage of DEAP was more prevalent among emotion studies, their findings \emph{do not} explicitly reflect emotion of music due to DEAP obtained the data from watching and listening to music videos, not only focused on music listening \cite{lin_24}. Therefore, studying preference and familiarity from listening to music, including measuring EEG, is an open topic of interest.}



\textcolor{black}{Alarcão and Fonseca showed that EEG and music preference had a plenty of rooms for investigation than emotion \cite{alarcao_13}. Only a few studies focused on classifying preference by analyzing only EEG \cite{hadjidimitriou_26}, fNIRS \cite{yamada_27}, and applied the audio features to EEG for better classification \cite{sawata_28}.
Hadjidimitriou \textit{et al.} classified the preference using EEG and demonstrated that the familiar music was able to provide better classification performance than unfamiliar music \cite{hadjidimitriou_25}.
Furthermore, they compared the time to initial EEG response between the best result of familiar and unfamiliar music, which unveiled that the familiar music required less time than the unfamiliar music. Their time result correlated with \cite{bachorik_28}, which investigated time to initial response to the assessment after listening to music. Unquestionably, preference, familiarity, response rate, and EEG showed a promising correlation between each other. However, to our best knowledge, none has examined all of these factors together.}


In order to investigate the relationship between these various factors, we constructed a database that includes preference, familiarity, response time (response rate), and EEG. In addition, these factors were investigated the distinction of participants' perception from listening to two music types: music without lyrics (Melody) and music with lyrics (Song). Both music types were an identical section, which was the \textit{first chorus section}. In this study, three main contributions are presented as follows:

\begin{enumerate}

\item \textcolor{black}{
We conducted the statistical analysis to investigate the distinction of music preference through familiarity, response rate, and EEG. We then examined the brain's appropriate region (left-right hemisphere and frontal) to optimize the music preference classification.}

\item \textcolor{black}{
\textcolor{red}{We utilized a support vector machine, random forest, and k-nearest neighbors for music preference classification} to observe the relevance to the statistical analysis of EEG. Then, familiarity and response rate were supplemented to the classifications for improving the performance, aiming to prove the influence of them, which should be considered when a music emotion experiment is designed.}


\item \textcolor{black}{We released the data of this study to the public, aiming to contribute to the study of music preference, familiarity, response rate, and EEG, as well as different music types (Melody and Song). We divided the music into 4 preference groups for analysis, which were favored Melody, favored Song, non-favored Melody, and non-favored Song. The dataset is available at https://github.com/IoBT-VISTEC/MUSEC-MusicPreference-Behaviors-EEG).}


\end{enumerate}

\textcolor{black}{
The remainder of this study is structured as follows: Section II describes materials and methods; Section III describes the recording and the pre-processing of the EEG; Section IV is results and data analysis; Section V is discussion; and the last section, Section VI, is for conclusion.}

\section{Methods and Materials}
\textcolor{black}{
In this section, we described the participants, music stimuli selection, detail of factors that we analyzed, and experimental protocol. The overview of this study showed in \Cref{experimental_framework}. }
\subsection{Participants}
\textcolor{black}{Twenty healthy participants (10 males, 10 females), whose native language is Thai, with no report of hearing impairment and brain diseases, aged from 22 to 35 (mean: 25.75, SD: 2.88), participated in the experiment. None of them had taken any musical lesson in recent years. They received a payment of 12.5 USD for attending the experiment. All the participants have signed a consent form before the experiment, and the study was approved by King Mongkut's University of Technology Thonburi Ethics Committee (KMUTT-IRB-COA-2020-027).}

\subsection{Music stimuli selection}


\textcolor{black}{
We utilized the Top 200 weekly chart of Thailand from renowned music streaming as Spotify \cite{spotify}. The music chart was ranked by the number of weekly streaming. 
We chose the top 20 music of each week and excluded redundant and international music. Finally, we had 110 popular Thai music from 31/08/2017 to 07/09/2019 (106 weeks).}

\textcolor{black}{
In this study, the music stimuli were divided into two music types: music without lyrics (Melody) and music with lyrics (Song). An identical section, \textit{first chorus section}, was used in both Melody and Song. The length were 19-66 seconds approximately.}





\subsubsection{Melody} \textcolor{black}{The Melody was transcribed 
by a musical expert as a classical pianist, who passed Yamaha piano performance fundamental examination grade 5 and Trinity Associate of Trinity College London. The Melody contained melodic pattern of the first chorus section and chords in the first beat of each chord and bar, which was written as MusicXML by MuseScore3 \cite{musescore}. The MusicXML was then rendered to .mp3 by Grand piano sound from GarageBand \cite{garageband}, in which the file quality was 44.1 kHz. }

\subsubsection{Song} \textcolor{black}{We purchased the Song from iTunes and trimmed the first chorus section using GarageBand. Fading about 0.5 seconds was included among the starting and the ending to avoid an unexpected situation for the participants when the Song started. The Song was rendered to .mp3, 44.1 kHz, and stereo.
}

\begin{figure*}
    \centering
    \includegraphics[width=1.7\columnwidth]{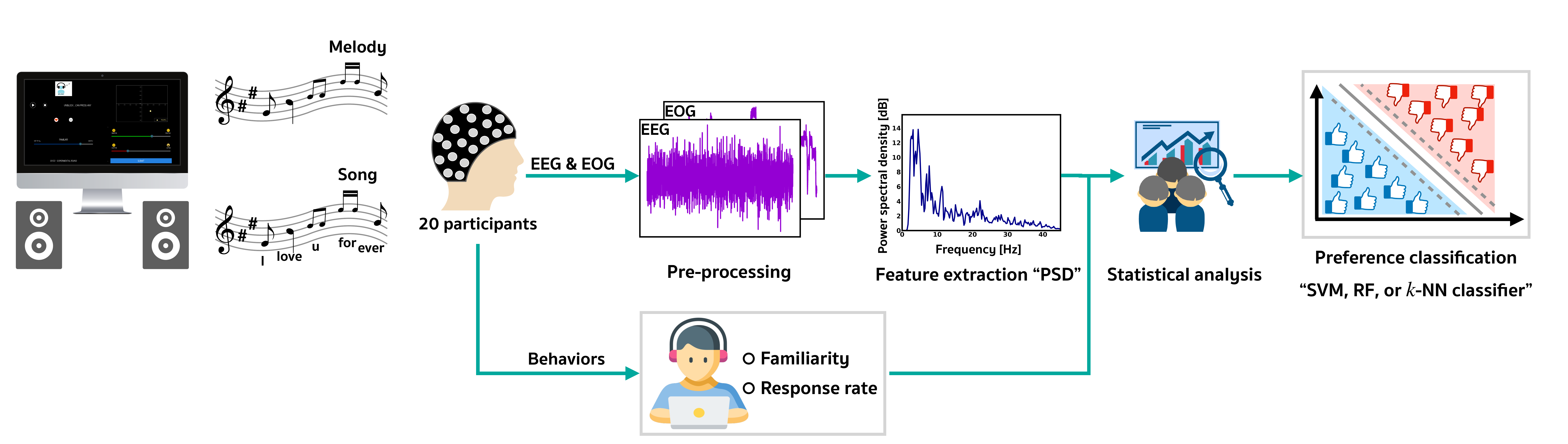}
    \caption{\textcolor{black}{A overview of this study.}}
    \label{experimental_framework}
\end{figure*}
\section{EEG Recording and Preprocessing}

\begin{figure}
    \centering
    \includegraphics[width=\columnwidth]{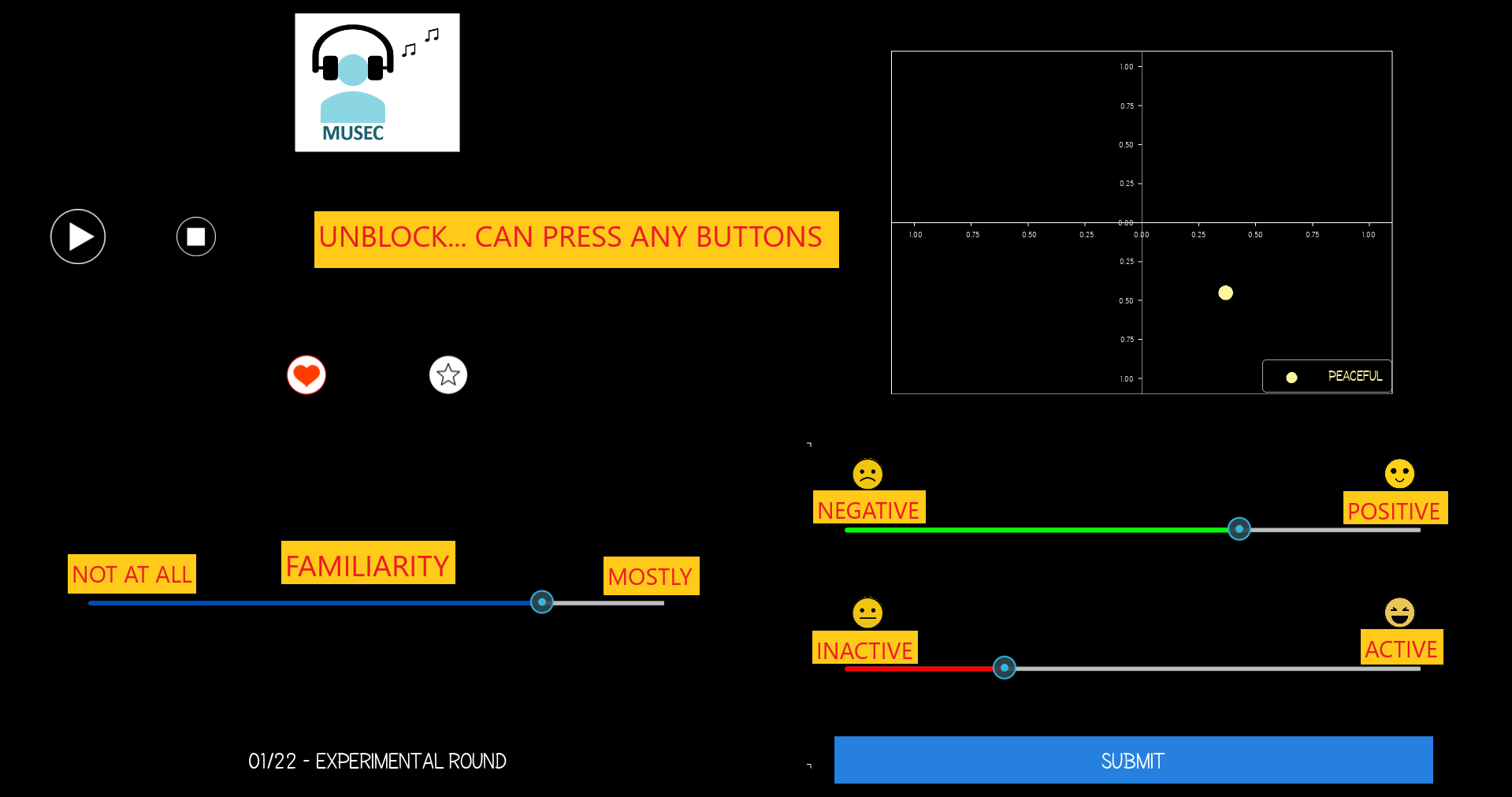}
    \caption{\textcolor{black}{Graphic user interface (GUI) of this study - The screen displayed the message as 'Unblock'. Participant was allowed to do self-assessment in Assessment-round. For example, the participant scored the familiarity tended to high (mostly), valence tended to high (positive), arousal tended to low (inactive), and participant favored this music.}}
    \label{musec_gui_scoring}
\end{figure}

\begin{figure}
    \centering
    \includegraphics[width=\columnwidth]{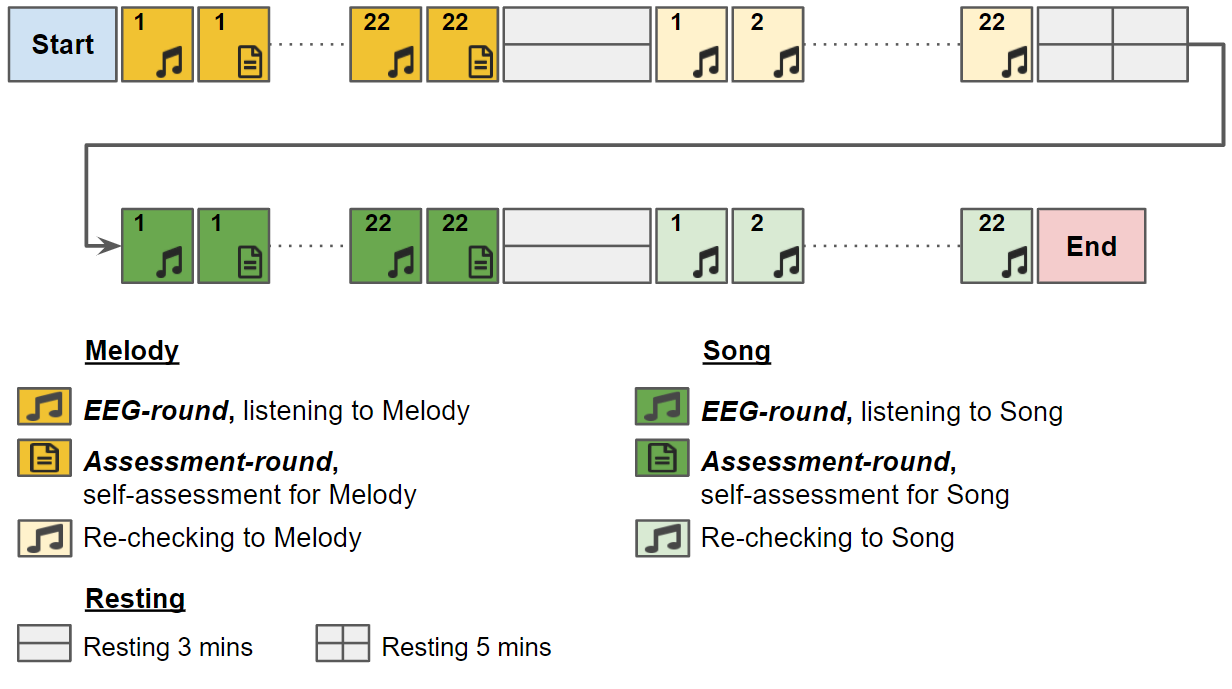}
    \caption{\textcolor{black}{Experimental protocol of this study. We investigated data from EEG-round and Assessment-round (dark yellow for Melody and dark green for Song). However, both re-checking rounds were not included in this study's scope (soft yellow for Melody and soft green for Song).}}
    \label{experimental_protocol}
\end{figure}


\subsection{Music preference selection}
\textcolor{black}{
As shown in \Cref{musec_gui_scoring}, we developed a individual software for the experiment. The experimental software used for collecting ground-truth data of music preference from participants. They used a heart button for reacting to favored music, which was up to their willing whether to react or not. In this study, music preference was divided into 2 groups such as \textit{favored} and \textit{non-favored}.}

\textcolor{black}{
Music were marked as the favored if participants pressed the heart button in both Melody and Song.
On the contrary, music was marked as the non-favored if the heart button was \textit{not} pressed in both Melody and Song. Meanwhile, music that had been favored in only Melody or Song were categorized as undecided, which was not counted and used in this study, since it could not be interpreted as either the favored or the non-favored.}





\subsection{Behaviors of Music Preference}
\textcolor{black}{
To investigate the distinction of music preference, we investigated the familiarity and response rate, which were defined as \emph{behaviors} as follows:}
\subsubsection{Familiarity (Fam)} \textcolor{black}{The participants expressed their familiarity with each music stimuli via a slider.
The range of slider between -1 and 1 was represented on the experimental software by “not at all” and “mostly”, respectively. The participants were told to use this function. They could not listen to next music stimuli if they did not use it.}

\subsubsection{Response Rate (ResR)} \textcolor{black}{The response rate reflected how fast the participants assess  each music stimuli. We secretly counted the time from the start until the end of the Assessment-Round (Describing the details in Section II-E) without informing the participants.
The response rate was calculated from counted time using the following equation:}

\begin{equation}
\label{response_rate_equation}
ResR (Hz) = \frac{1}{Time (seconds)}
\end{equation}

\textcolor{black}{Taking note that the valence (negative-positive), arousal (inactive-active), and adding to playlist (star button), were also recorded in the experiment, however, the recorded data were not used in this study.}

\subsection{Experimental Protocol}
\textcolor{black}{To avoid misinterpretation, an instructor methodically presented the usage of software's functions and an experimental protocol in \Cref{experimental_protocol} to the participants. Then, 3 practice trials were served to participants to familiarize themselves with the experimental software, in which the practice music stimuli were not used in the main experiment. Participants could ask the instructor all the time if encountered with any problem before the practice trials ended. } 

\textcolor{black}{
Then, we divided the 20 participants into 4 groups of 5 participants. For each group, the 110 music stimuli were randomly divided into 5 playlists of 22 music stimuli. Therefore, all participants did not listen to the redundant playlist.
Methodically, participants were asked to listen to the 22 Melody first, followed by the 22 Song.}

\textcolor{black}{Music stimuli were played via an Ultimate Ears Boom Bluetooth speaker. The music volume was allowed to adjust via the experimental software before the experiment started. During the experiment, participants were not able to adjust the volume.
The experimental protocol was as follows:}


\subsubsection{EEG-round}: \textcolor{black}{The participants were able to only click the play button to play the music stimuli. After clicking, 5-second silence were presented for baseline, and then a music stimulus was automatically played. Participants were not allowed to use any software's functions until the music stimulus had ended because \emph{Brain response was examined in this EEG-round.}}

\subsubsection{Assessment-round}: \textcolor{black}{the screen displayed a message that allow the participants to do self-assessment using software's functions. We secretly counted the response rate (1) 
in this round from the start of self-assessment until the participants press the submit button. After the Assessment-round, the participants went to next music stimuli until completed all 22 music stimuli.}

\textcolor{black}{After listening to 22 Melody, the participants took a 3-minute break. Then, they went through to its re-checking round.
Next, they took a 5-minute break, and then listening to 22 Song, which had the same round pattern as Melody. However, taking note that both re-checking rounds were not included in this study's scope.
(soft yellow for Melody and soft green for Song in \Cref{experimental_protocol}). We only utilized the self-assessment data from both Assessment-round. Finally, the experiment used of approximately one and a half hour.}



\textcolor{black}{In this experiment, a biosignal amplifier (g.USBamp RESEARCH, g.tec, Austria) was used to simultaneously measured EEG and EOG signals with the sampling rate of 1200 Hz. The EEG data were acquired by using 62 active electrodes and placed according to the 10-20 international system. The EEG channels were referenced and grounded using the right earlobe and the $Fpz$ electrode, respectively. The EOG signals were obtained from two electrodes positioned under and next to the outer canthus of the right eye. The impedance of both EEG and EOG electrodes were maintained below 10 k$\Omega$ during the entire experiment.}

\textcolor{black}{For the preprocessing, the notch filter at 50 Hz was applied to remove the electrical noises. Then, recorded EEG signals were high-pass filtered at 0.1 Hz using 5\textsuperscript{th} order non-causal Butterworth filter. To remove the artifacts in the EEG data, including eye blinks, this paper applied an eye movement-related artifact correction-based on independence component analysis (ICA)\cite{fastICA}. 
The recorded EEG data was referenced to Common Average Reference (CAR) for improving spatial resolution \cite{Chella_2016}. Furthermore, the recorded EEG signals were band-pass filtered between 2-45 Hz, using 4\textsuperscript{th} order non-causal Butterworth filter.}

\begin{table}[t]
    \caption{List of channels for EEG analysis.}
    \centering
    \label{eeg_channels}
    \resizebox{\linewidth}{!}{
        \begin{tabular}{@{}lc@{}}
        \toprule[0.2em]
        \textbf{Regions}            & \textbf{Channels}          \\ \midrule[0.1em] \midrule[0.1em] 
        Hemisphere\_Left & Fp1, AF3-5-7, F1-3-5-7, FT7, FC1-3-5, T7,\\
        & C1-3-5, TP7, CP1-3-5, P1-3-5-7, PO3-7, O1 \\ \midrule[0.1em]
        Hemisphere\_Right &  Fp2, AF4-6-8, F2-4-6-8, FT8, FC2-4-6, T8,\\
        & C2-4-6, TP8, CP2-4-6, P2-4-6-8, PO4-8, O2  \\ \midrule[0.1em]
        Frontal\_Left & Fp1, AF3-7, F1-3-5-7, FT7, FC1-3-5\\ \midrule[0.1em]
        Frontal\_Right & Fp2, AF4-8, F2-4-6-8, FT8, FC2-4-6 \\
        \bottomrule[0.2em]
        \end{tabular}
        }
\end{table}

\section{Results and data analysis}
\textcolor{black}{
Ground-truth of music preference contained the favored (N = 81) and the non-favored (N = 271). We statistically scrutinized the distinction of music preference through familiarity, response rate and EEG, which defined significance as p $<$ 0.05 using Mann-Whitney test. Subsequently, the statistical analysis were utilized as the features for music preference classification by support vector machine, random forest and k-nearest neighbors as shown in \Cref{experimental_framework}.}

\begin{figure*}
\centering
  \subfloat[\label{familiarity_fig}]{%
       \includegraphics[width=.48\linewidth]{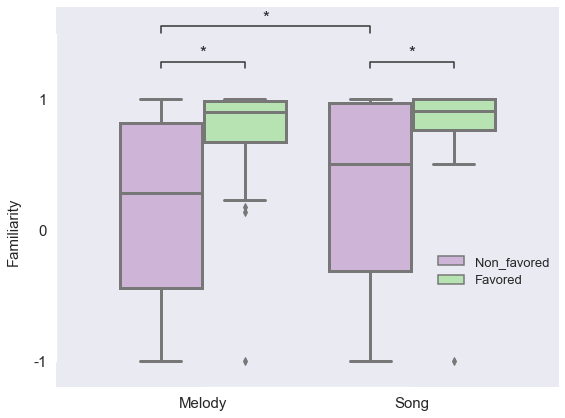}}
    \hspace*{\fill}   
  \subfloat[\label{response_rate_fig}]{%
        \includegraphics[width=.48\linewidth]{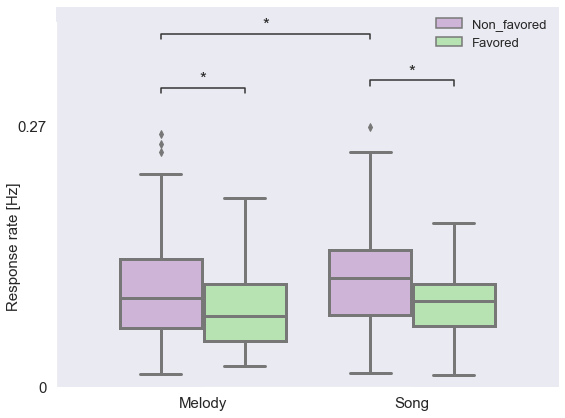}}
  \caption{\textcolor{black}{Pairwise comparisons of (a) familiarity and (b) response rate. Significance was defined as p $<$ 0.05.}}
   \label{familiarity_responsera} 
\end{figure*}

\begin{figure*}
\centering
  \subfloat[Melody\label{correlation_Melody}]{%
       \includegraphics[width=.48\linewidth]{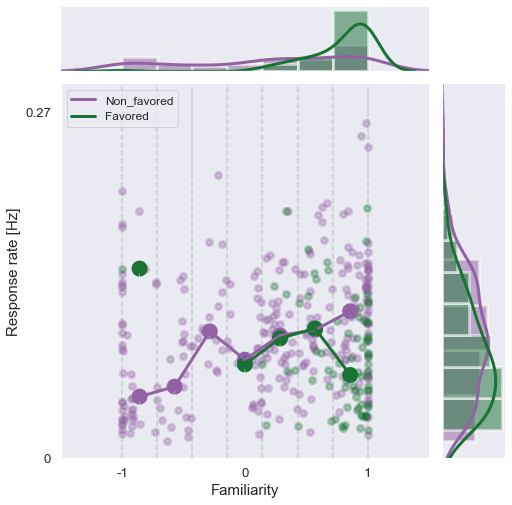}}
    \hspace*{\fill}   
  \subfloat[Song\label{correlation_Song}]{%
        \includegraphics[width=.48\linewidth]{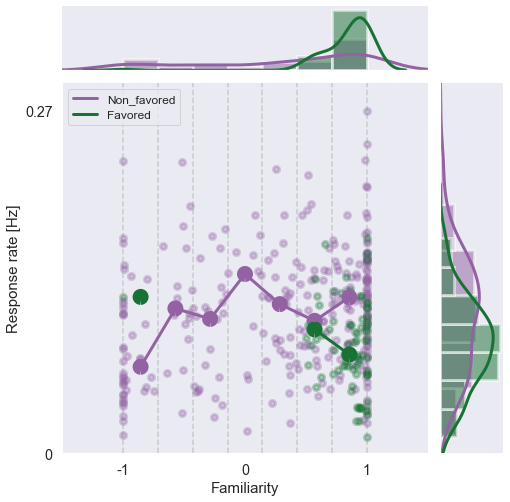}}

  \caption{\textcolor{black}{Correlation between familiarity and response rate using median trend line. (a) Melody and (b) Song.}}
\end{figure*}

\begin{figure*}
\centering
  \subfloat[Melody\label{eeg_hemisphere}]{%
       \includegraphics[width=.48\linewidth]{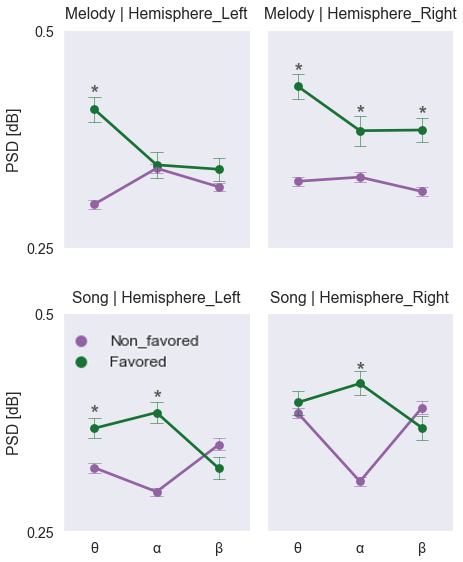}}
    \hspace*{\fill}   
  \subfloat[Song\label{eeg_frontal}]{%
        \includegraphics[width=.48\linewidth]{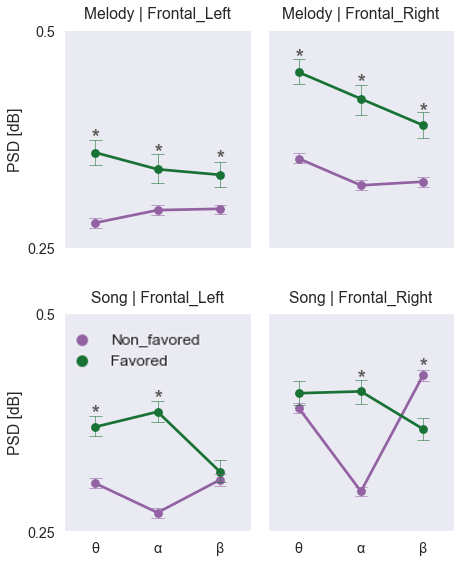}}

  \caption{\textcolor{black}{Power spectral density average with standard error of favored and non-favored music, which illustrated the 3 frequency bands as theta ($\theta$), alpha ($\alpha$), and beta ($\beta$) of left and right hemisphere from listening to (a) Melody and (b) Song. Significance was defined as p $<$ 0.05 for pairwise comparison between the favored and the non-favored.}}
\end{figure*}

\subsection{Familiarity (Fam)}

\textcolor{black}{
As shown in \Cref{familiarity_fig}, the analysis showed that the lack of lyrics did not affect the result of favored music. Melody alone was sufficient for the participants to recognize their favored music, confirming by high familiarity ($>$ 0) in both Melody and Song, which had no difference. On the contrary, although the medians of both non-favored Melody and Song gravitated towards high familiarity, but we found data distributed to low familiarity ($<$ 0) with significance (p $<$ 0.05). Likewise, the distinction of music preference in its Melody and its Song were explicitly illustrated the significance (p $<$ 0.05) for both.}



\subsection{Response Rate (ResR)}
\textcolor{black}{
We calculated the response rate with (1).
As shown in \Cref{response_rate_fig}. The significance appeared to be similar to the familiarity’s result, including the significance of the distinction of music preference in its Melody and Song (p < 0.05), the significance of non-favored Melody and Song (p < 0.05), and no significance of favored Melody and Song. Thus, in the favored, which had high familiarity, the response rate showed that the participants spent more time than the non-favored in both Melody and Song.}

\subsection{Correlation between familiarity and response rate}
\textcolor{black}{
The results of the familiarity and response rate seemed to be a positive correlation. Therefore, we investigated the correlation between these factors using a median trend line of Melody and Song, which were displayed in \Cref{correlation_Melody} and \Cref{correlation_Song}, respectively. For Melody, in high familiarity ($>$ 0), the favored and the non-favored showed the identical tendency, which was high familiarity required less time. However, in the highest familiarity, which had the most data, we found the distinction of music preference (p $<$ 0.05). This meant the participants explicitly gravitated towards spending more time on the favored than the non-favored when listening to the highest familiar music. For Song, we discovered this distinction and significance (p $<$ 0.05) in the highest familiarity, similar to Melody.}

\subsection{Power spectral density (PSD)}
\textcolor{black}{
Debating about left and right of music's brain response is ongoing. Albouy \textit{et al.} reported that left side was for processing a speech content, while right side was for a music content \cite{albouy_34}. Therefore, to study the distinction of brain response to Melody and Song, we scrutinized the big regions of the brain, which were left and right hemisphere. Then, the investigation was scoped to the left and right frontal. Because the frontal brain was observed and mentioned in the EEG-based emotion from listening to stimulus\cite{lin_24, geethanjali_32, lee_33}.}

\textcolor{black}{
We calculated power spectral density (PSD) of left-right hemisphere and left-right frontal from listening to Melody and Song, which were filtered into theta (4-7 Hz), alpha (8-13 Hz), and beta (13-30 Hz) bands and normalized. The channels of each hemisphere and frontal were presented in \Cref{eeg_channels}. We reported the PSD's average of each hemisphere and frontal in \Cref{eeg_hemisphere} and \Cref{eeg_frontal}, respectively.}


\subsubsection{Left-Right hemisphere}
\textcolor{black}{In \Cref{eeg_hemisphere}, for Melody, the distinction of music preference showed that right hemisphere clarified over left hemisphere. We found significance in all bands from right hemisphere but only in the theta band from left hemisphere.}


\textcolor{black}{For Song, we found the significance in theta and alpha bands from left hemisphere, and only in alpha band from right hemisphere. However, Song did not clarify the distinction between left and right hemisphere as Melody. Also, the distinction of music preference was not clear compared to that from Melody in right hemisphere.}            


\subsubsection{Left-Right frontal}
\textcolor{black}{
In \Cref{eeg_frontal}, for Melody, the left frontal showed the significance in all bands, which differed from that of left hemisphere. Meanwhile, the significance was found in all bands from right frontal, similar to right hemisphere.}


\textcolor{black}{
For Song, the significance of theta and alpha were found in left frontal, similar to left hemisphere. Meanwhile, the difference between right frontal and right hemisphere was in beta band's significance, which appeared in frontal only.}

From the results, significance appeared in all bands of both Melody's right hemisphere and frontal. Then, we hypothesized that music preference classification in case of Melody using PSD from right side likely outperform the left side.




\subsection{Music preference classification}
Support vector machine (SVM), Random forest (RF), and K-Nearest neighbors (kNN) were utilized to prove the feasibility of the statistical analysis.
GridSearchCV tuned all models' hyperparameters \textcolor{red}{(\Cref{hyperparameters} in Appendix)} with Stratified sampling 10-fold cross-validation. During training, we used one fold as a testing set, one fold as a validation set, and the rest as a training set. For an evaluation, we focused on the F1-score due to imbalance classes. This section reported and emphasized highest F1-scores in {\Cref{model_results}} \textcolor{red}{(Accuracy showed in Appendix)}.




\subsubsection{Without behaviors}

\textcolor{red}
{For Melody, the highest F1-score of SVM and kNN obtained from the right frontal feature, including 53.18\% and 48.73\%, respectively. On the contrary, RF obtained from the left hemisphere, which highly contrasted with our assumption. However, RF could not outperform the SVM and kNN, which were the models that showed correlated results with the statistical analysis.}

\textcolor{red}
{For Song, as shown in \Cref{eeg_hemisphere} and \Cref{eeg_frontal}, the statistical distinction between the left and right side did not become evident like Melody. Therefore, the brain region that performed the highest F1-score was ambiguous.
}




\subsubsection{With behaviors}
\textcolor{red}{We were able to clarify the importance of factors, whether adopting the familiarity (\emph{Fam}), response rate (\emph{ResR}), or both familiarity and response rate (\emph{Fam + ResR}). All results with these behaviors showed in supplementary, Table VI-VIII.}

\textcolor{red}{
For Melody, as shown in \Cref{bestf1_withbehaviors}, all highest F1-scores resulted from using both familiarity and response rate. SVM with the right frontal still achieved the highest F1-score (71.09 \%), similar to the result of without behaviors.}

\textcolor{red}{
For Song, it was similar to Melody, in which using both familiarity and response rate achieved the highest F1-scores of SVM and kNN, except RF model that achieved from using familiarity only. Meanwhile, most highest F1-scores of both Melody and Song still resulted from the right side.}

\section{Discussion}



\begin{table}

    \caption{\textcolor{red}{Our highest F1-scores from SVM, RF, and kNN. (a) Without behaviors and (b) With behaviors. The brain region that performed the best F1-score are defined as \emph{fr}: Right frontal, \emph{hr}: Right hemisphere, \emph{hl}: Left hemisphere, \emph{flr}: Left-Right frontal, and \emph{hlr}: Left-Right hemisphere.}}
    \centering
    \subfloat[Without behaviors\label{bestf1_withoutbehaviors}]{%
        \begin{tabular}{@{}lcc@{}}
            \toprule[0.2em]
            \textbf{Models} & \textbf{Melody} & \textbf{Song} \\ \midrule[0.1em] \midrule[0.1em]
            SVM & \textbf{53.18} $\pm\ \textbf{4.87}^{fr}$ & 47.53 $\pm\ 3.95^{hr}$  \\ 
            RF & 46.27 $\pm\ 1.62^{hl}$ & 50.85 $\pm\ 2.77^{flr}$  \\ 
            kNN & 48.73 $\pm\ 2.25^{fr}$ & \textbf{51.91} $\pm\ \textbf{2.55}^{hlr}$  \\ 

            \bottomrule[0.2em]
        \end{tabular}}
        
    \subfloat[With behaviors (\textit{Fam, ResR)}\label{bestf1_withbehaviors}]{%
        \begin{tabular}{@{}lcc@{}}
            \toprule[0.2em]
            \textbf{Models} & \textbf{Melody} & \textbf{Song} \\ \midrule[0.1em] \midrule[0.1em]
            SVM & \textbf{71.09} $\pm\ \textbf{2.70}^{fr}$ & 58.67 $\pm\ 6.38^{fr}$  \\ 
            RF & 66.57 $\pm\ 3.22^{hl}$ & \textbf{74.92} $\pm\ \textbf{2.80}^{hr}$  \\
            kNN & 67.40 $\pm\ 3.40^{hr}$ & 68.77 $\pm\ 2.90^{fr}$  \\ 
            \bottomrule[0.2em]
        \end{tabular}}
    \label{model_results}
\vspace{-6mm}
\end{table}

\begin{table*}
\centering
\caption{\textcolor{red}{Studies of music preference classification. (\emph{pm}: proposed method)}}
\label{tab:comparison}
\begin{tabular}{@{}ccccccc@{}}
\toprule[0.2em]
\textbf{Ref.} &
\textbf{Participants} &
\begin{tabular}[c]{@{}c@{}}\textbf{Music Stimuli}\\ \textbf{(length)}\end{tabular} & \textbf{Features} &
  \begin{tabular}[c]{@{}c@{}}\textbf{Analyzed brain region/} \\ \textbf{No. of channels (channels)}\end{tabular} &
  \textbf{Classifier} &
  \textbf{Highest Accuracy / F1-score} ($\pm \textbf{SE}$) \\ \midrule[0.1em] \midrule[0.1em]
{\cite{sawata_28}} &
  10 &
  \begin{tabular}[c]{@{}c@{}}60 music\\ (15 s)\end{tabular} &
  EEG, Audio &
  \begin{tabular}[c]{@{}c@{}}All regions/ \\ 12 Channels (Fp1-2, \\ F7-8, C3-4, P3-4, O1-2, T3-4)\end{tabular} &
  SVM &
  81.40 $\pm\ 6.00$ / 81.70 $\pm\ 5.00$ \\ \midrule[0.1em]
{\cite{bajoulvand_50}} &
  16 &
  \begin{tabular}[c]{@{}c@{}}4 folk music\\ (120 s)\end{tabular} &
  EEG &
  \begin{tabular}[c]{@{}c@{}} Frontal/\\ 4 Channels (Fp1-2, F7-8)\end{tabular} &
  SVM &
  83.20 $\pm\ 5.50$ / 62.90 $\pm\ 0.30$ \\ \midrule[0.1em]
\begin{tabular}[c]{@{}c@{}} Melody$^{pm}$\end{tabular} &
  \textbf{20} &
  \begin{tabular}[c]{@{}c@{}}\textbf{110} Melody\\ (19-66 s)\end{tabular} &
  EEG, Behaviors &
  \begin{tabular}[c]{@{}c@{}}Right frontal/ \\ 11 Channels (see \Cref{eeg_channels})\end{tabular} &
  \begin{tabular}[c]{@{}c@{}}SVM \end{tabular} &
  75.29 $\pm\ 2.45$ / 71.09 $\pm\ 2.70$ \\ \midrule[0.1em]
 \begin{tabular}[c]{@{}c@{}}Song$^{pm}$\end{tabular} &
  \textbf{20} &
  \begin{tabular}[c]{@{}c@{}}\textbf{110} Song\\ (19-66 s)\end{tabular} &
  EEG, Behaviors &
  \begin{tabular}[c]{@{}c@{}}Right hemisphere/ \\ 27 Channels (see \Cref{eeg_channels})\end{tabular} &
  \begin{tabular}[c]{@{}c@{}}RF \end{tabular} &
  84.64 $\pm\ 1.59$ / 74.92 $\pm\ 2.80$ \\
  
  \midrule[0.2em]
\end{tabular}
\end{table*}

\textcolor{black}{
The purpose of this study is to study the influential factors for eliciting the innermost music emotion from listening to the first chorus section without lyrics (Melody) and with lyrics (Song). Music preference is a primary influential factor, which was observed and reported in previous studies. We then examined the distinction of music preference through familiarity, response time (response rate), and brain response.}



\subsection{Familiarity}
\textcolor{black}{
As shown in \Cref{familiarity_fig}, we confirmed the positive correlation between music preference and familiarity reported in \cite{north_20, hamlen_20}. The favored clarified the high familiarity over the non-favored in both Melody and Song. 
We implied that participants were able to recognize the favored even if listening to the melodic pattern only. On the contrary, the melodic pattern of the non-favored was not sufficient to recognize,
in spite of listening to the most memorable section as the first chorus section.
Therefore, reaching to top 20 of the music chart did not confirm the recognition or familiarity with the music.}

\textcolor{black}{
We concluded that the favored and non-favored can be classified by familiarity. Psychologically, the result could also interpret the correlation with a \textit{mere-exposure effect} \cite{madison_47, peretz_47, meyer_48}. However, how preference has an impact on familiarity, and vice versa, as well as their influence on music emotion are still interesting open questions \cite{canon_6}.
}

\subsection{Response Rate}
\textcolor{black}{ We included an observation of time spent in self-assessment for each music stimuli by calculating response rate using (1).
The result unveiled that the favored spent more time than the non-favored in both Melody and Song. We may interpret that participants tend to regard the self-assessment in the favored over the non-favored, despite they had high familiarity with both favored Melody and Song. According to the results, we are able to indirectly hypothesize that the familiarity and the response rate were the positive correlation.  However, the results seem to contrast with \cite{hadjidimitriou_25, bachorik_28}, which reported that familiar music was spent less time than the unfamiliar. Thus, we directly investigated the correlation between familiarity and response rate for finding the reason of contradiction.}

\subsection{Correlation between familiarity and response rate}
\textcolor{black}{
As shown in \Cref{correlation_Melody} and \Cref{correlation_Song}, in the highest familiarity which had the most data, both Melody and Song showed the distinction of music preference. Participants spent more time on the favored. Therefore, our result was not completely contrast with \cite{hadjidimitriou_25, bachorik_28}. We agree with them in case of our non-favored music, which spent less time in familiar music. Interestingly, this finding confirms that studying the music preference is essential. We can explain the contradiction, which lead to insight of human.}

\textcolor{black}{
In case we hypothesize that spending more time means mindfully doing the self-assessment, which is able to elicit the innermost music emotion. The results may approve the previous works, whether preference and familiarity were the influential factors for music emotion study. 
However, spending more time is whether good or not is an interesting question.}

\textcolor{black}{
From the results, we clarify the distinction of preference
from familiarity and response rate. To summarize, the favored has very high familiarity, spending more time in both Melody and Song. Meanwhile, the non-favored is on the contrary. }

\subsection{Power spectral density (PSD)}
\textcolor{black}{
We statistically scrutinized power spectral density of left-right hemisphere and frontal, as shown in \Cref{eeg_hemisphere} and \Cref{eeg_frontal}, respectively. Listening to the melodic pattern without lyrics (Melody) would clarified that the right side outperformed over the left side. Meanwhile, ambiguous distinctions of the left-right occurred in listening to Song, which has both speech (lyrics) and music contents. These results may conform to the assumption that left and right sides processed speech and music content, respectively \cite{albouy_34}. } 

\textcolor{black}{
For real world application, we executed music preference classification using machine learning models (SVM, RF, and kNN) with features from the statistical PSD analysis. Then, we addressed the familiarity and response rate to the models. We observed the highest F1-score, as shown in \Cref{model_results}. }


\subsection{Music preference classification}
\textcolor{red}{
In both without and with behaviors, we found that the right sides appeared on most of the highest F1-scores, which related to our assumption that the right side outperformed that of the left, especially the right frontal (\emph{fr}).}

\textcolor{red}{
In most cases, using both familiarity and response rate increased the F1-scores, except the RF model on Song using only familiarity.
Meanwhile, we found several of Song’s F1-scores decreased only using the response rate (see Table VI-VIII in supplementary). Although these behaviors could describe the distinction of preference by statistical analysis, we could not conclude the correlation between statistical behaviors and all models. Therefore, executing a feature selection should be a concern.}

\textcolor{red}{We compare our highest F1-scores of Melody (71.09\%) and Song (74.92\%) with other studies in \Cref{tab:comparison}. Our proposed behaviors potentially improve the F1-score to the same level as the other studies. We can outperform \cite{bajoulvand_50}, but not \cite{sawata_28}. This comparison confirms that these behaviors (familiarity, and response rate) are worth investigating in underlying music research, including, preference, emotion and others. 
}
\textcolor{red}{
However, it is complicated to directly compare because of many differences between the studies, including analyzed brain regions/channels, feature vectors, machine learning models, numbers of participants, and especially music stimuli.
}

\subsection{Music stimuli selection}
As shown in \Cref{experimental_framework}, music stimuli are a starting point of such an experiment, which were used in the form of short excerpt. Unquestionably, different sections of its music can induce different perceptions, which lead to misinterpretation and inaccurate annotation. Similar to the feature selection, regarding music stimuli selection is important \cite{warrenburg_15}.

\textcolor{black}{
In this study, participants listened to the \textit{first chorus section} since the section is the most memorable part of the popular (Thai) music. Also, the other chorus sections had a chance to encounter the distinct harmony or melody for soloing or singing with modulation, which has a chance to induce different perceptions.}
\textcolor{black}{
Besides, \cite{hult_17} showed that using the chorus section dataset as PMEmo \cite{zhang_18} with ground truth from a highly controlled environment performed better than the large dataset with ground truth from social tags on Last.fm \cite{song_19} and crowd-sourcing \cite{aljanaki_19}. They suggested that the high quality dataset might be more required than the large annotated dataset. As a result, we supported that the music stimuli selection should be carefully concerned. It is a influential factor to the music perception study.}




\textcolor{black}{
However, behind music stimuli contain many music theories. We are aware that our selection is not an accurate selection. Using chorus section could not apply to the other genres, which have complicated theories and feelings, such as classical or jazz music. Thus, investigation of music theory and perception is an ongoing question for investigating what appropriate section and length of music excerpt are. }

\subsection{Leveraging the experiment for industry}
\textcolor{black}{
This study contribute the experimental design of music emotion toward music industry.
Experiment with music preference, familiarity, and response rate can be easily executed in underlying study. 
However, to investigate further study with brain response, high-priced equipment is a major issue. Suggestively, penetration of consumer grade EEG measuring sensors become an optional solution due to its lower price \cite{sawangjai_40}. In emotion study, consumer grade EEG measuring sensors became a part of investigation \cite{lakhan_19, katsigiannis_49,sml21}. Nevertheless, the gap between medical and consumer grade EEG measuring sensors have a plenty of room for research. Fulfilling this gap will increase the performance in collecting listener's data in daily life for driving the music industry to the next level.}

\textcolor{black}{
Finally, we believe that music has a potential impact on the contribution to many industries, not only to music industries. We suggest applying the experiment to biomedical and healthcare industries since preferred and familiar music are commonly mentioned in studying the autism \cite{damota_44, greenberg_45, greenberg_46} and dementia \cite{bartfay_41, ward_42, ekra_43}. In fact, understanding musical taste of other people is complicated matter, even for those who are close. Thus, the experiment in this paper has an ability to collect the musical taste from the autism and mild cognitive impairment (the risk of dementia). Hopefully, our experimental design can be utilized to improve the efficiency of music therapy in the future.}

\section{Conclusion}
\textcolor{black}{
People tend to be more familiar and spend more time assessing their favored music than non-favored music, no matter listening to Melody or Song. The result can interpret that participants carefully do their self-assessment of the favored, which can elicit music's innermost emotion. \textcolor{red}{In classifying the preference using statistical analysis on EEG, we found that the right side provides the higher F1-scores than the left side, especially the right frontal, which agrees with our assumption.} By supplementing the familiarity and response rate to the classification, we ultimately achieved enhanced F1-scores. Finally, we can clarify the distinction of music preference by using statistics and machine learning models. The results can contribute to the experimental design of music emotion. In future work, we will collect more data from participants, who are native and non-native language, aiming to observe a cross-cultural music cognition. Lastly, we will investigate the emotion of music between the favored and the non-favored, proving that these factors are influential in eliciting the innermost emotion.}


\section{Appendix}
\begin{table}[ht]
    \caption{\textcolor{black}{List of hyperparameters tuned for SVM, RF, and kNN by using GridSearchCV.}}
    \label{hyperparameters}
    \centering
        \begin{tabular}{@{}lcc@{}}
            \toprule[0.2em]
            \textbf{Models} & \textbf{Parameter (kernel)} & \textbf{Values} \\
            \midrule[0.1em] \midrule[0.1em]
            SVM & $C$ (All) & 0.001, 0.01, 0.1, 1, 10, 100 \\ 
            & $\gamma$ (All) & 0.001, 0.01, 0.11, 1, 10 \\ 
            & $d$ (Poly)  & 2, 3, 4\\ \midrule[0.1em]
            RF & No. of estimators  & 100, 300, 500\\
            & Max depth  & 5-17 with a step of 3\\
            & Min sample leaf  & 2-14 with a step of 3\\
            & Min sample split  & 2-14 with a step of 3\\
            \midrule[0.1em]
            kNN & No. of neighbors  & 10-49\\
            & Weights  & Uniform, Distance\\
            & Metric  & Euclidean, Manhattan, Minskowski\\
 
            \bottomrule[0.2em]
        \end{tabular}
\end{table}

\begin{table}[ht]
    \caption{Our highest accuracy from SVM, RF, and kNN. (a) Without behaviors and (b) With behaviors. The brain region that performed the highest F1-score are defined as \emph{fr}: Right frontal, \emph{hr}: Right hemisphere, \emph{hl}: Left hemisphere, \emph{flr}: Left-Right frontal, and \emph{hlr}: Left-Right hemisphere.}
    \centering
    \subfloat[Without behaviors\label{bestaccuracy_withoutbehaviors}]{%
        \begin{tabular}{@{}lcc@{}}
            \toprule[0.2em]
            \textbf{Models} & \textbf{Melody} & \textbf{Song} \\ \midrule[0.1em] \midrule[0.1em]
            SVM & 67.75 $\pm\ 5.35^{fr}$ & 70.30 $\pm\ 5.25^{hr}$  \\ 
            RF & 76.14 $\pm\ 0.69^{hl}$ & 77.85 $\pm\ 1.11^{flr}$  \\ 
            kNN & 77.56 $\pm\ 0.82^{fr}$ & 78.42 $\pm\ 0.79^{hlr}$  \\ 

            \bottomrule[0.2em]
        \end{tabular}}
        
    \subfloat[With behaviors (\textit{Fam, ResR)} \label{bestaccuracy_withbehaviors}]{%
        \begin{tabular}{@{}lcc@{}}
            \toprule[0.2em]
            \textbf{Models} & \textbf{Melody} & \textbf{Song} \\ \midrule[0.1em] \midrule[0.1em]
            SVM & 75.29 $\pm\ 2.45^{fr}$ & 71.13 $\pm\ 5.55^{fr}$  \\ 
            RF & 80.08 $\pm\ 1.83^{hl}$ & 84.64 $\pm\ 1.59^{hr}$  \\ 
            kNN & 77.80 $\pm\ 2.15^{hr}$ & 80.68 $\pm\ 1.49^{fr}$  \\ 
            \bottomrule[0.2em]
        \end{tabular}}
\vspace{-6mm}
\end{table}

\bibliographystyle{IEEEtran}
\bibliography{Reference}

\end{document}